\documentclass[twocolumn,showpacs]{revtex4}
\usepackage{amssymb}
\usepackage{graphicx}
\usepackage{dcolumn}
\usepackage{amsmath}

\begin{document}
\preprint{HEP/123-qed}
\title{Adsorption and two-body recombination of atomic hydrogen on $^3$He-$^4$He mixture films}
\author{A.I. Safonov,$^{1,2}$ S.A. Vasilyev,$^1$ A.A. Kharitonov,$^2$ S.T. Boldarev,$%
^{1,3}$ I.I. Lukashevich,$^2$ and S. Jaakkola$^1$}
\address{$^1$Wihuri Physical Laboratory, Department of Physics, University of Turku, 20014 Turku,
Finland\\
\\
$^2$Laboratory of Metastable Quantum Systems, ISSSPH, Kurchatov Institute, 123182 Moscow, Russia\\
\\
$^3$P.L.Kapitza Institute for Physical Problems, RAS, 117334 Moscow, Russia}
\date{\today}
\begin{abstract}
We present the first systematic measurement of the binding energy $E_a$ of hydrogen atoms to the
surface of saturated $^3$He-$^4$He mixture films. $E_a$ is found to decrease almost linearly from
1.14(1) K down to 0.39(1) K, when the population of the ground surface state of $^3$He grows from
zero to $6\times10^{14}$ cm$^{-2}$, yielding the value $1.2(1)\times 10^{-15}$ K~cm$^2$ for the
mean-field parameter of H-$^3$He interaction in 2D. The experiments were carried out with overall
$^3$He concentrations ranging from 0.1 ppm to 5 \% as well as with commercial and isotopically
purified $^4$He at temperatures 70...400 mK. Measuring by ESR the rate constants $K_{aa}$ and
$K_{ab}$ for second-order recombination of hydrogen atoms in hyperfine states $a$ and $b$ we find
the ratio $K_{ab}/K_{aa}$ to be independent of the $^3$He content and to grow with temperature.
\end{abstract}
\maketitle

Two-dimensional (2D) Bose systems acquire growing interest since the observation of local
coherence in the weakly interacting gas of hydrogen atoms adsorbed on liquid helium surface
\cite{saf98}. We found that at high quantum degeneracy the probability of three-body surface
recombination of H atoms is suppressed at least by a factor of 11(2). The phenomenon was
attributed to the formation of a 2D quasicondensate (QC), condensate with fluctuating phase.
However, for an ideal gas three-body recombination in the condensate would be a factor of 3! less
probable than in the noncondensate \cite{KSS87} and H-H interactions are expected to make the
effect of QC on local correlations even smaller \cite{svist99}. The difference between the
experimental and theoretical suppression factors has been accounted for on the basis of increasing
delocalization of the bound state wave function in surface-normal direction with increasing
density of the 2D hydrogen \cite{kagan89,stoof}. Thus it seems desirable to experimentally
separate the roles of local coherence and delocalization. Experiments with 2D hydrogen on
$^3$He-$^4$He mixtures are anticipated to give such an opportunity by virtue of weaker \cite{blue}
and, as it will be shown here, even tunable binding of H atoms to the surface.

In this paper we report on experiments where the binding energy $E_a$ of hydrogen to liquid helium
has been measured as a function of $^3$He surface coverage. $E_a$ is found to decrease almost
linearly from 1.14(1) K to 0.39(1) K, when the $^3$He coverage grows from zero to about one atomic
layer. We have also studied two-body recombination of H atoms in their two lower hyperfine states
$a$ and $b$, since the probability of this process does not change upon the appearance of QC and
thus serves as reference for three-body recombination \cite{saf98}. Our data unambiguously
corroborates the prediction \cite{greben} that the recombination rate constant ratio
$K_{ab}/K_{aa}$ increases with temperature.

In our experiments ESR operating at 128 GHz (field $B$ = 4.57 T) and NMR at 910 MHz have been
employed, respectively, to measure and control the $a$ and $b$ state populations. The versatile
combination of ESR and NMR provides well-defined conditions for studying different recombination
processes. Moreover, the data analysis is simple and reliable. The ESR spectrometer has been
calibrated calorimetrically with an estimated absolute accuracy of 10\% and long-term stability of
2\%. The minimum detectable density is about $2\times10^{12}$ cm$^{-3}.$

The volume of H gas in the sample cell, including the annexed ESR and NMR cavities, is $V$ = 4.5
cm$^3$. The area of inner cell walls coated with helium film is $A$ = 22 cm$^2$. The cell
temperature $T$ is measured with a RuO$_2$ thick-film resistor attached to the outer cell wall and
calibrated against the $^3$He melting curve with an absolute accuracy of 1 mK.

\begin{figure}[tbp]
\includegraphics[width=8.5cm]{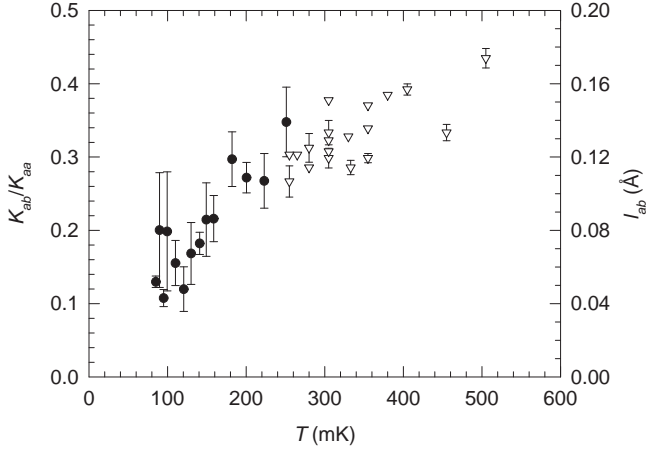}
\caption{Temperature dependence of the rate constant ratio $K_{ab}/K_{aa}$ (left scale) and the
$ab$ recombination crosslength $l_{ab}$ (right scale) obtained in this work ($\bullet$) and by
Statt~\textit{et al.} \cite{statt} ($\triangledown$).}
\label{fig:gamma}
\end{figure}

At the relatively low densities considered here the H gas both in 2D and in 3D is well described
by Boltzmann statistics. The recombination of hydrogen atoms takes place in the adsorbed phase,
whereas the large majority of atoms is in the bulk. Then the effective rate constant of the
second-order decay of the bulk density is related to the intrinsic rate constant of two-body
surface recombination by the relation \cite{blue}
\begin{equation}
K_{ij}^{\mathrm{eff}}={\frac AV}\lambda^2\exp{\ \left({\frac{2E_a}{kT}}\right)}K_{ij},
\label{eq:Keff}
\end{equation}
where $\lambda =\sqrt{2\pi \hbar ^2/mkT}$ is the thermal de~Broglie wavelength, $m$ is the
hydrogen atomic mass and subscripts $i$ and $j$ denote the hyperfine states. Therefore, by
measuring the temperature dependence of $K_{ij}^{\mathrm{eff}}$ one can determine both $E_a$ and
$K_{ij}$ as has been done for H on $^4$He in many previous experiments \cite{blue}. In case of
equally populated states $a$ and $b$ one has
\begin{equation}
{\frac{dn_b}{dt}}={\frac{dn_a}{dt}}=-(K_{aa}^{\mathrm{eff}}+K_{ab}^{\mathrm{eff}})n_b^2.
\label{eq:dn/dt}
\end{equation}

We apply rf power to the sample saturating the $b\leftrightarrow a$ NMR transition to continuously
equalize the two populations. Being to a high accuracy linear in time \cite{R}, the measured
$1/n_b$ gives $K^{\mathrm{eff}}\equiv K_{aa}^{\mathrm{eff}}+K_{ab}^{\mathrm{eff}}$. Evolution of
$1/n_a$ observed in a separate experiment yields exactly the same values of $K^{\mathrm{eff}}$. In
an experiment of another type short rf pulses resonant with the $b\rightarrow a$ transition are
used to convert a small fraction of the otherwise $b$-state sample to the $a$-state. From the
lifetime $\tau=1/K_{ab}^{\mathrm{eff}}n_b$ of $a$-atoms we obtain $K_{ab}^{\mathrm{eff}}$
\cite{jetp}. We believe that this procedure to determine $K_{aa}^{\mathrm{eff}}$ and
$K_{ab}^{\mathrm{eff}}$ is more reliable than extraction of the rate constants from multiparameter
non-linear fits of the density decays to coupled rate equations.

\begin{figure}[t]
\includegraphics[width=8.5cm]{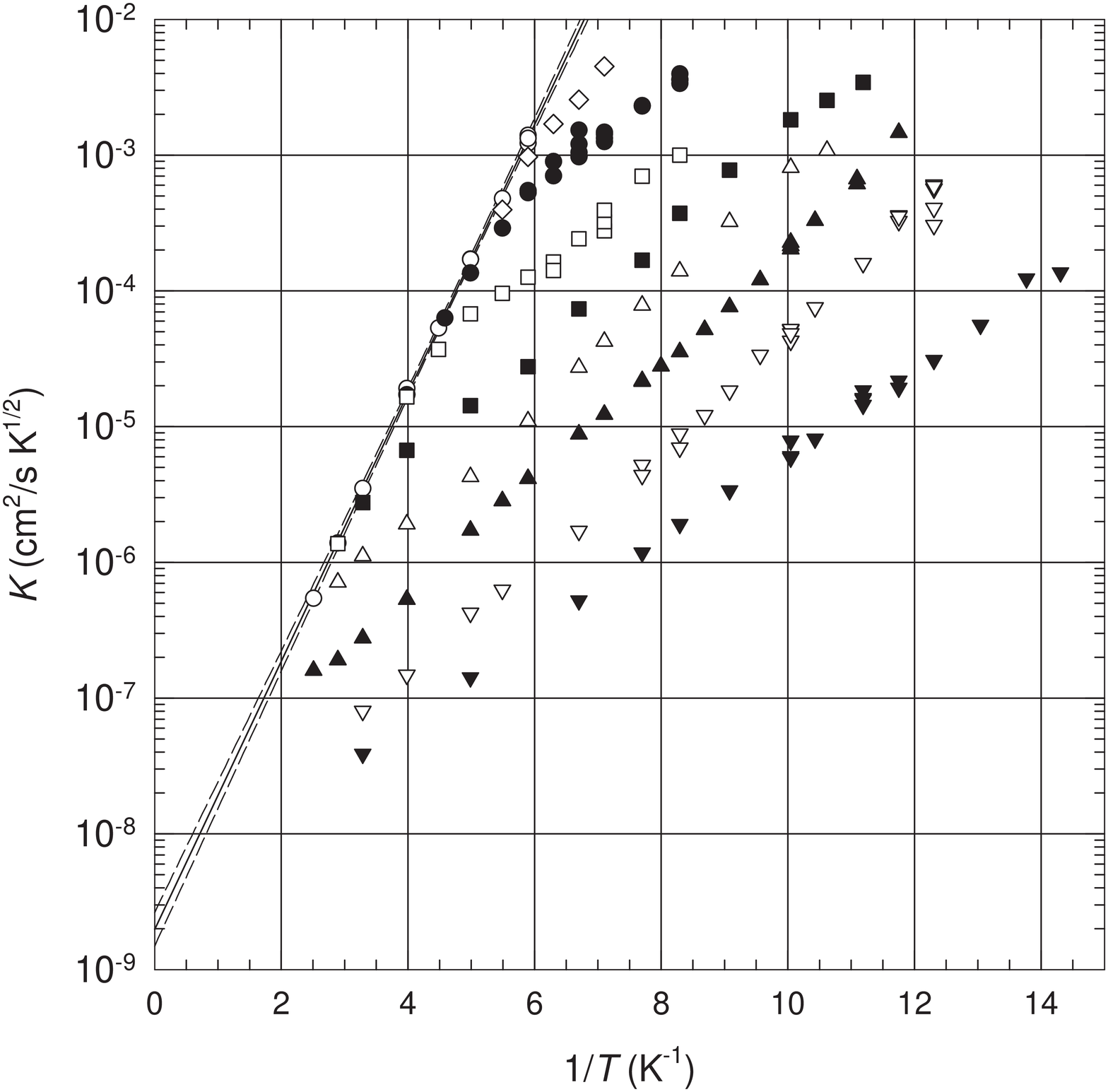}
\caption{The effective rate constant plotted as $K$ =
$\frac{V}{A}K^{\mathrm{eff}}_{aa}/\lambda^2\protect\sqrt{T}$ versus $1/T$ for H on isotopically
pure ($\circ$) and commercial ($\diamondsuit$) $^4$He, 0.1 ($ \bullet$), 1 ($\square$), 10
($\blacksquare$), and 100 ($\vartriangle$) ppm, 0.1 ($\blacktriangle$), 1 ($\triangledown$), and 5
($\blacktriangledown$) \% of $^3$He. Solid and dashed lines represent linear regression and its
99\% confidence intervals, respectively, for pure $^4$He.}
\label{fig:Keff}
\end{figure}

The experiments are carried out at $T$ = 70...400 mK for overall $^3$He concentrations $c_3$ =
0.1, 1, 10, and 100 ppm, 0.1, 1, and 5\% as well as for isotopically purified $^4$He ($\lesssim$ 1
ppb of $^3$He) and commercial helium. The total amount of liquid helium in the cell is about 11
cm$^3$. The free surface area accessible to $^3$He is 57 cm$^2$ including the film-covered
low-temperature part of the H inlet tube coming from a dissiciator.

Fig.~\ref{fig:gamma} shows the temperature dependence of the ratio
$K_{ab}^{\mathrm{eff}}/K_{aa}^{\mathrm{eff}}$ obtained in this work together with the results of
Statt~\textit{et al.} \cite{statt}. In agreement with the theory of Greben~\textit{et al.}
\cite{greben}, but in contradiction with some earlier experimental results \cite{blue}, the rate
constant ratio grows with $T$. In fact, $ab$ recombination produces only ortho-H$_2$ with odd
angular momentum and consequently the atoms must overcome a centrifugal barrier \cite{statt}. The
probability of such a process vanishes at $T$ = 0. We also emphasize that, within experimental
scatter, $K_{ab}^{\mathrm{eff}}/K_{aa}^{\mathrm{eff}}$ shows no systematic change upon addition of
$^3$He and is therefore averaged over all concentrations. The error bars in Fig.~\ref {fig:gamma}
represent standard deviation of the data.

In Fig.~\ref{fig:Keff} the temperature dependence of $K_{aa}^{\mathrm{eff}}$ is presented for
various $^3$He concentrations and for commercial as well as isotopically pure $^4$He. Following
Ref.~\onlinecite{greben}, we assume the crosslength $l_{aa}$ for $aa$ surface recombination to be
temperature independent. Then $K_{aa}=\overline{v}l_{aa}\varepsilon ^2\propto \sqrt{T}$, where
$\varepsilon \simeq 2.53\times 10^{-2}$T/$B$ is the hyperfine mixing parameter and
$\overline{v}=\sqrt{\pi kT/m}$ the relative thermal velocity in 2D. For H on pure $^4$He $E_a$
does not vary with $T$. Then half of the slope of the
$\ln{(\frac{V}{A}K^{\mathrm{eff}}_{aa}/\lambda^2\protect\sqrt{T})}$ versus $1/T$ line is
$E_a=1.14(1)$ K and the intercept gives $l_{aa}=0.40(10) \mathrm{\AA}$. For $T$-independent
$l_{aa}$ Fig.~\ref {fig:gamma} may also be regarded as the temperature variation of the
crosslength $l_{ab}$.

\begin{figure}[t]
\includegraphics[width=8.5cm]{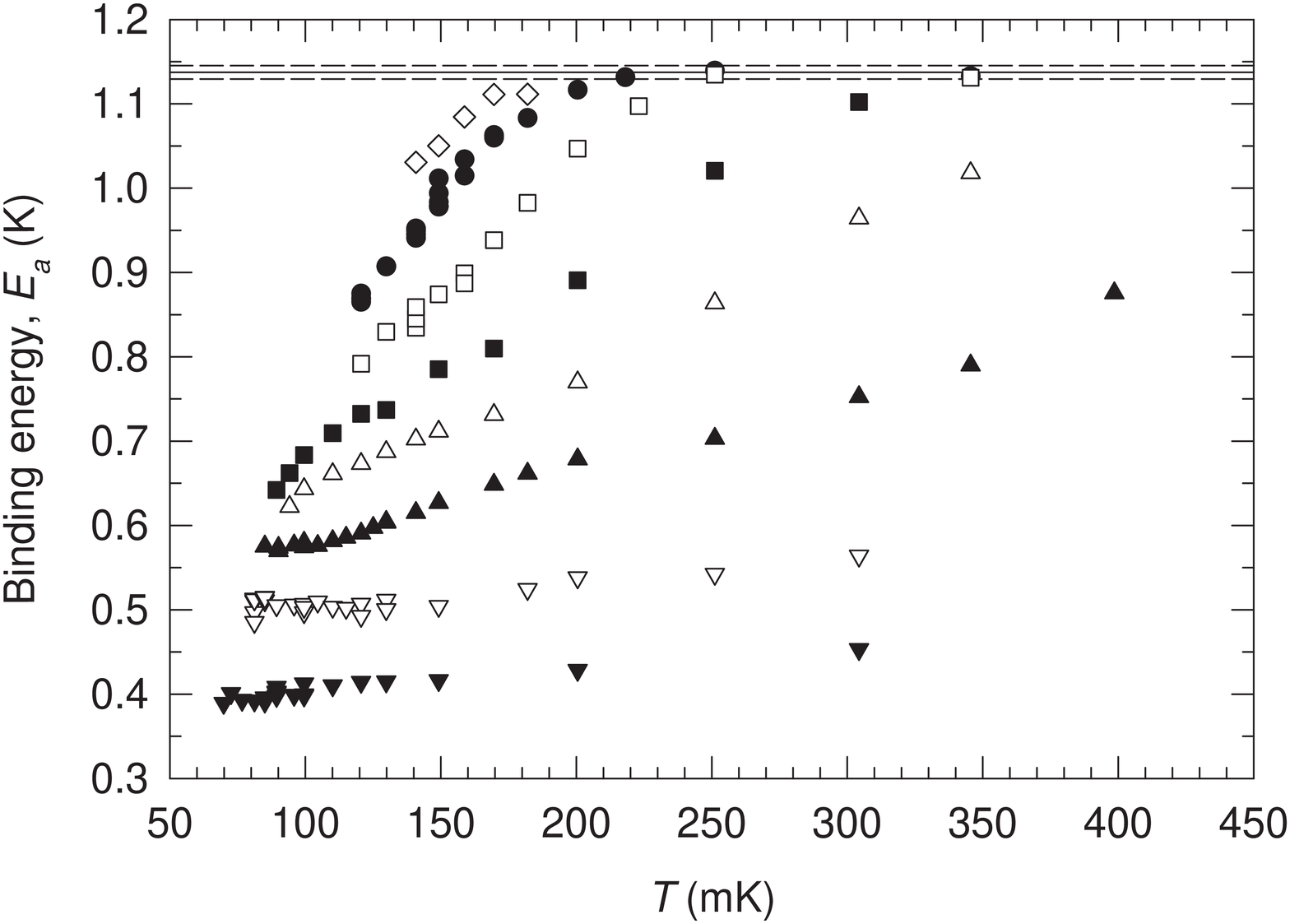}
\caption{Temperature dependence of $E_a$ for the same $^3$He concentrations as in
Fig.~\ref{fig:Keff}. The lines are the average value and confidence intervals for $E_a$ of H on
pure $^4$He.}
\label{fig:Ea-T}
\end{figure}

In Fig.~\ref{fig:Keff} the data also for $c_3=5\%$ fall on a straight line. Yet the $^3$He surface
coverage changes \cite{ES} and we cannot take $E_a$ to be constant. Instead we assume, to the
first approximation, that the crosslength $l_{aa}$ does not depend on $^3$He content either. Then
we determine $E_a$ for each concentration and temperature from the measured
$K_{aa}^{\mathrm{eff}}$ using Eq.~(\ref{eq:Keff}) with $l_{aa}$ fixed to its pure-$^4$He value.
The results are shown in Fig.~\ref {fig:Ea-T}, from which one notices that the behavior of $E_a$
resembles that of the surface tension of $^3$He-$^4$He solutions \cite{ES}. Both have their origin
in 2D $^3$He bound to the surface of the liquid. Since the chemical potential of H atom inside
liquid helium is large positive, even a small overlap of the wavefunctions of adsorbed H and
$^3$He repels hydrogen atoms from the surface, i.e., raises the energy level of the single bound
state of the H atom.

According to Pavloff and Treiner \cite{PT} there are at least two surface states of $^3$He on bulk
$^4$He. The binding energies relative to the bulk liquid and the effective masses in zero coverage
limit are $e_{s0}=2.64$ K and $M_0=1.29m_3$ for the ground state and $e_{s1}=0.81$ K and
$M_1=1.6m_3$ for the excited state. Here $m_3$ denotes the bare mass of $^3$He. The variations of
these quantities with $^3$He coverage are also given in Ref.~\onlinecite{PT}. In good agreement
with Ref.~\onlinecite {PT} the occupation of the excited surface state at $n_{3s}\geq 3.5\times
10^{14}$ cm$^{-2}$ has been recently observed in an experiment \cite{cool}. Therefore, using the
above values we may calculate the populations $n_{3s0}$ and $n_{3s1}$ of the both $^3$He surface
states for all concentrations $c_3$ and temperatures.

\begin{figure}[t]
\includegraphics[width=8.5cm]{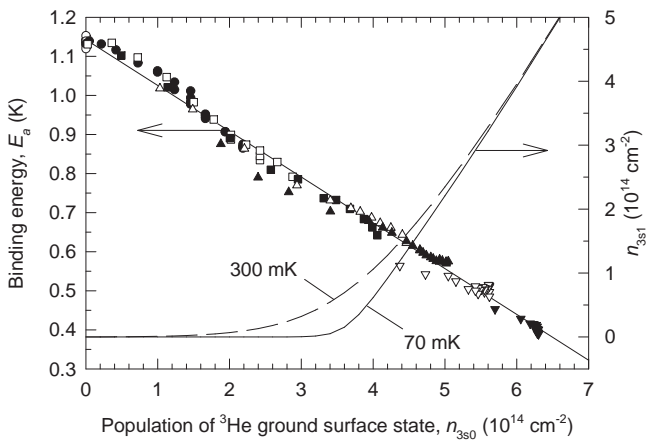}
\caption{Dependence of H binding energy on $^3$He surface density for the same concentrations as
in Figs.~\ref{fig:Keff} and \ref{fig:Ea-T}. Curved lines are populations of the excited surface
state of $^3$He calculated for $T$ = 70 and 300 mK.}
\label{fig:Ea-n3}
\end{figure}

Fig.~\ref{fig:Ea-n3} presents the hydrogen adsorption energy $E_a$ as a function of $n_{3s0}$. The
population of the excited state is also shown for reference. The decrease of $E_a$ is obviously
due to interaction of adsorbed H atoms with the surface states of $^3$He. The slope of the $E_a$
vs. $n_{3s0}$ line is the effective mean-field parameter $U_{30}=1.2(1)\times 10^{-15}$ K~cm$^2$.
It does not seem to change at $n_{3s0}=3.5\times 10^{14}$ cm$^{-2}$, where the occupation of the
excited state begins. This points to the interaction of H with the excited state of $^3$He being
weak, $U_{31}\ll U_{30}$.

Despite of the above assumption we can consider two reasons why the crosslength $l_{aa}$ might
change upon addition of $^3$He. First, helium atoms play the role of the third body in hydrogen
recombination. Clearly, the probability of collisions with helium increases in presence of 2D
$^3$He due to a larger overlap of the wave functions. On the other hand, the surface-normal
delocalization length $d=\hbar/\sqrt{2E_am}$ of the hydrogen wavefunction grows with $n_{3s0}$.
The latter reason causes the H-H recombination crosslength $l_{aa}\propto 1/d\propto\sqrt{E_a}$
\cite{KSS87}, to decrease with growing $n_{3s0}$. Even if the two effects do not completely cancel
each other, our results for $E_a$ remain practically unchanged because the variation of $l_{aa}$
is important when both $T$ and $n_{3s0}$ are high, i.e., for very few data points only (cp.
Figs.~\ref{fig:Keff} and \ref{fig:Ea-n3}).

Confiding in our techniques to monitor and manipulate the hyperfine level populations in our H
samples we believe that the present work sheds new light on the long-standing discrepancies
between numerous previous determinations of $E_a$. Unknown hyperfine polarization is one of the
several possible sources of systematic error in the determination of $E_a$ listed by
Godfried~\textit{et al.} \cite{harvard}. Another source is $^3$He impurity and from
Fig.~\ref{fig:Keff} it is obvious that even a very small amount of $^3$He can considerably change
the average slope typically taken as $2E_a$. This would also affect the value of the recombination
crosslength $l_{aa}$ extracted from the $T$-dependence of the effective rate constants under the
assumption of constant $E_a$. It is likely that in several earlier studies ''pure'' $^4$He was, if
not explicitly stated otherwise, just non-purified commercial helium with unknown $ ^3$He
contamination. This is not a problem at $T\gtrsim$ 200 mK (cp. Fig.~\ref{fig:Ea-T}) or if there is
no bulk helium in the sample cell but a saturated film only. In the latter case the area-to-volume
ratio is so large, about $10^5$ cm$^{-1}$, that the $^3$He surface coverage is at most $~10^{-4}$
monolayers even for $c_3$ = 1 ppm which is at least an order of magnitude larger than the natural
abundance.

The ESR results of Statt~\textit{et al.} \cite{statt} for $K_{aa}^{\mathrm{eff}}$ and
$K_{ab}^{\mathrm{eff}}$ at $T$ = 250...500 mK agree quite well with ours scaled to same magnetic
field and $A/V$ ratio. Morrow~\textit{et al.} \cite{morrow} measured the zero field NMR frequency
shift for H above 200 mK and found, as coupled fitting parameters, the binding energy $E_a$ =
1.15(5) K and the wall shift $\Delta _s$ = -49(2) kHz. Later Pollack~\textit{et al.} \cite
{pollack} observed simultaneously surface and bulk H atoms by NMR and directly found $\Delta _s$ =
43.2(10) kHz, as extrapolated to $B$ = 0. Shinkoda and Hardy \cite{shinkoda} used ESR for a direct
detection of H atoms at the surface of $^4$He. They had several cm$^3$ of liquid helium in the
sample cell and measured the apparent $E_a$ to increase from 0.75 K to 1.03 K when $T$ increased
from 70 to 140 mK. This observation is just an extrapolation of our results for commercial helium
(open diamonds in Fig.~\ref{fig:Ea-T}) to lower temperatures. Associating the results of Refs.
\onlinecite{statt,morrow,pollack,shinkoda}, our value $E_a$ = 1.14(1) K for H on isotopically pure
$^4$He and the theoretical prediction \cite{greben} with each other we gather strong support for
the assumption of a temperature independent crosslength $l_{aa}$. It should also be added that
another potential pitfall in the $E_a$ determination was avoided in this work by extending the
measurements over a wide enough temperature range.

It is worth comparing the lowest value of the binding energy, $E_a=0.39(1)$ K, measured in our
experiment for $c_3$ = 5\% with earlier results extracted from pressure measurements of density
decays. The latter are 0.39(1) K for pure $^3$He \cite{jochemsen} and 0.34(3) K for the mixture of
two parts of $^3$He and one part of $^4$He \cite{yperen}.

The linear decrease of $E_a$ with $n_{3s}$ shows that H-$^3$He interaction is well described by
the mean-field approximation. This is an important observation as such, but it seems to disagree
with the following arguments. Typically the Fermi energy of 2D $^3$He gas is much higher than
temperature, $e_{0F}=\pi \hbar ^2n_{3s0}/2M_0\gg kT$. Then elastic H-$^3$He collisions in 2D
responsible for decreasing $E_a$ involve $^3$He quasiparticles from the Fermi level only. Their
density, of order $n_{3s0}T/e_F=2M_0T/\pi \hbar ^2$, is independent of $n_{3s0}$. On the other
hand, the Fermi level rises with $n_{3s0}$, the corresponding wave function expands in
surface-normal direction \cite{PT} and overlaps more with adsorbed hydrogen.

The present measurements of hydrogen adsorption energy $E_a$ as a function of $^3$He surface
density offer a unique opportunity to perform experiments on a degenerate 2D Bose gas with tunable
interaction strength which scales as $\sqrt{E_a}$. As such, the quantum system of a degenerate 2D
Bose gas (H) coupled with a degenerate 2D Fermi gas ($^3$He) would also be an interesting object
to study. The mean-field parameter $U_{30}$ obtained here should play a key role in the behavior
of that system.

This work is supported by Academy of Finland, INTAS (grant No 97-11981), Wihuri Foundation, the
Russian Ministry of Industry, Science and Technology and RFBR (grant No 99-02-17357).


\end{document}